\documentclass[prb,reprint,superscriptaddress]{revtex4-1}

\usepackage[english]{babel}
\usepackage{graphicx}
\usepackage{xspace}
\usepackage{fancybox}
\usepackage{epic}
\usepackage{calc}
\usepackage{ifthen}
\usepackage{float}
\usepackage{color}
\usepackage{amsmath,amssymb}
\usepackage{rotating}
\usepackage{subfigure}
\usepackage{exscale}
\usepackage{setspace}
\usepackage[latin1]{inputenc}
\usepackage{longtable}

\DeclareMathAlphabet{\mathbfit}{OT1}{cmr}{bx}{it}

\begin{document}

\title{Magnetism and spin-orbit coupling in Ir-based double perovskites La$_{2-x}$Sr$_x$CoIrO$_6$}

\author{A.~Kolchinskaya}
\affiliation{Institute of Materials Science, Technische
Universit\"{a}t Darmstadt, Petersenstr.~23, 64287 Darmstadt, Germany}

\author{P.~Komissinskiy}
\affiliation{Institute of Materials Science, Technische
Universit\"{a}t Darmstadt, Petersenstr.~23, 64287 Darmstadt, Germany}

\author{M.~Baghaie Yazdi}
\affiliation{Institute of Materials Science, Technische
Universit\"{a}t Darmstadt, Petersenstr.~23, 64287 Darmstadt, Germany}

\author{M.~Vafaee}
\affiliation{Institute of Materials Science, Technische
Universit\"{a}t Darmstadt, Petersenstr.~23, 64287 Darmstadt, Germany}

\author{D.~Mikhailova}
\affiliation{Institute of Materials Science, Technische
Universit\"{a}t Darmstadt, Petersenstr.~23, 64287 Darmstadt, Germany}
\affiliation{Institute for Complex Materials, IFW Dresden, P.O. Box 270116, 01171 Dresden, Germany}

\author{N.~Narayanan}
\affiliation{Institute of Materials Science, Technische
Universit\"{a}t Darmstadt, Petersenstr.~23, 64287 Darmstadt, Germany}
\affiliation{Institute for Complex Materials, IFW Dresden, P.O. Box 270116, 01171 Dresden, Germany}

\author{H.~Ehrenberg}
\affiliation{Karlsruhe Institute of Technology (KIT), Inorganic Chemistry, Engesserstr.~15, D-76131 Karlsruhe, Germany}

\author{F.~Wilhelm}
\affiliation{ESRF, ID-12, 6 Rue Jules Horowitz, BP 220, 38043 Grenoble, France}

\author{A.~Rogalev}
\affiliation{European Synchrotron Radiation Facility, ID-12, 6 Rue Jules Horowitz, BP 220, 38043 Grenoble, France}

\author{L.~Alff}
\email{alff@oxide.tu-darmstadt.de}
\affiliation{Institute of Materials Science, Technische
Universit\"{a}t Darmstadt, Petersenstr.~23, 64287 Darmstadt, Germany}

\date{10 October 2011}

\begin{abstract}

We have studied Ir spin and orbital magnetic moments in the double perovskites La$_{2-x}$Sr$_x$CoIrO$_6$ by x-ray magnetic circular dichroism. In La$_2$CoIrO$_6$, Ir$^{4+}$ couples antiferromagnetically to the weak ferromagnetic moment of the canted Co$^{2+}$ sublattice and shows an unusually large negative total magnetic moment (-0.38\,$\mu_{\text B}$/f.u.) combined with strong spin-orbit interaction. In contrast, in Sr$_2$CoIrO$_6$, Ir$^{5+}$ has a paramagnetic moment with almost no orbital contribution. A simple kinetic-energy-driven mechanism including spin-orbit coupling explains why Ir is susceptible to the induction of substantial magnetic moments in the double perovskite structure.

\end{abstract}
\maketitle

\section{\textbf{INTRODUCTION}}

Double perovskites of the form $A_2BB'$O$_6$, with $A$ an earth alkaline metal, and $B$ and $B'$ a $d$ transition metal, have attracted considerable attention in recent years. Within this class of materials, there are compounds with properties such as a high Curie temperature, $T_{\text{C}}$,\cite{Kobayashi:98,Kato:02a,Krockenberger:07,Serrate:07} a high magnetoresistance,\cite{Kobayashi:98} a metal-insulator transition \cite{Aligia:01,Kato:02b,Poddar:04}, and half-metals.\cite{Philipp:03} This huge variety of properties has its origin in the possibility of doping and substituting the perovskite structure at the $A$ and $B$ sites, allowing tailoring of the electronic, crystal, and magnetic structure of the compounds which, in turn, interact with each other.
Sr$_{2}$FeMoO$_{6}$  was the first double perovskite for which a high magnetoresistance at room temperature was reported ($T_{\text{C}}>420$ K).\cite{Kobayashi:98} By electron doping in similar compounds the Curie-temperatures rise to 635\,K in Sr$_{2}$CrReO$_{6}$ \cite{Kato:02a,Majewski:05a} and even up to 750 K in Sr$_{2}$CrOsO$_{6}$,\cite {Krockenberger:07,Das:11} which is so far the highest Curie temperature observed in ferrimagnetic double perovskites. In previous measurements, a kinetic-energy-driven exchange model, where ferromagnetism is stabilized by hybridization between the magnetic and the non-magnetic/weakly magnetic ions, has been well confirmed.\cite{Sarma:00,Philipp:03} This hybridization-driven mechanism is in competition with simple superexchange.\cite{Das:11}

The most important key to finding novel materials with increased Curie temperatures is the understanding of the magnetic coupling of the $B$ and $B'$ ions. Considering the existing compounds, it is obvious that the combination of a strongly magnetic ion and a typically non-magnetic or weakly magnetic ion such as Mo, Ru, W, Re, and Os, may result in double perovskite ferrimagnets with extraordinarily high Curie temperatures.  Thus, the understanding of magnetic coupling will reveal routes to designing improved materials. Here, we investigate Ir as the weakly magnetic element in an antiferromagnetic double perovskite, La$_{2-x}$Sr$_x$CoIrO$_6$.\cite{Narayanan:10} Despite the low $T_{\text{C}}$ values of these compounds between 70 and 95 K, they are a good study object to learn about the magnetic coupling of $3d$ elements to $5d$ transition metals with strong spin-orbit interaction.\cite{Lezaic:11} Furthermore, the experimental determination of site-specific magnetic moments offers the possibility of testing the prediction power and limitations of band structure calculations. For some compounds, the theoretical predictions are surprisingly close to the experimental results \cite{Majewski:05a,Majewski:05b}. However, most likely due to electronic correlations, phenomena remain which are difficult to understand theoretically, as, for example, the unusually high Curie temperature and metal-insulator transition in Ca$_2$FeReO$_6$.\cite{Westerburg:02,Kato:02b,Iwasawa:05,Winkler:09}

\section{\textbf{EXPERIMENT}}

Polycrystalline samples of La$_{2-x}$Sr$_x$CoIrO$_6$ with  $0\leq x\leq2$ were prepared by solid state synthesis. These samples have been characterized by x-ray powder diffraction, neutron powder diffraction (NPD), superconducting quantum interference device (SQUID) magnetometry, and synchrotron powder diffraction.\cite{Narayanan:10} The structural and magnetic Rietveld refinement made for the NPD measurements using FullProf \cite{FullPROF} reveal that for $x = 0, 0.5, 1, \text{and\,} 2$, the Co lattice orders antiferromagnetically, but with different types of antiferromagnetism (see below).\cite{Narayanan:10} However, the Ir magnetic moments cannot be refined from the NPD data. In order to investigate the magnetic coupling of the $3d$ Co ions and $5d$ Ir ions, element specific methods such as x-ray magnetic circular dichroism (XMCD) are mandatory. Samples of La$_{2-x}$Sr$_x$CoIrO$_6$ with $x = 0, 0.5, 1, \text{and\,} 2$ were measured at the beamline ID-12\cite{Rogalev:01} at the European Synchrotron Radiation Facility. Spectra were recorded using the total fluorescence yield detection mode. XMCD spectra were obtained as the
direct difference between consecutive X-ray absorption
near-edge spectroscopy (XANES) scans recorded with opposite helicities of the
incoming x-ray beam in 17 T at low temperature for the Co $K$ edge and the Ir $L_{2,3}$ edges.
The x-ray absorption spectra for right and left circularly polarized beams were corrected for self-absorption effects, taking into account the chemical composition, the density, an infinite thickness (justified by the sample thickness), the background contributions from the fluorescence of subshells and matrix as well as from coherent and incoherent scattering, the angle of incidence of the x-ray beam, and, finally, the solid angle of the detector.\cite{Goulon:82} The self-absorption corrections can be used safely since they have been proven to work extremely well in the case of U multilayers, where the self-absorption corrections are huge.\cite{Wilhelm:07} In our case, the difference between spectra corrected for self-absorption effects and as-measured spectra is at most 6\% at the maximum of the white line intensity for all samples. This can be understood by the fact that Ir is rather diluted in the matrix. The Ir $L_{3,2}$ edge-jump intensity ratio $L_3$/$L_2$ was then normalized to 2.22. \cite{Henke:93} This takes into account the difference in the radial matrix elements of the $2p_{1/2}$-to-$5d$($L_2$) and $2p_{3/2}$-to-$5d$($L_3$) transitions. A deviation of $\pm10$\% in the $L_3$/$L_2$ XAS edge-jump normalization would affect the branching ratio $B$ by $\pm2.5$\% and the moment analysis by $\pm5$\%.

\section{\textbf{RESULTS AND DISCUSSION}}

Previous measurements and Rietveld refinement of NPD results showed different types of magnetic order of Co for the La$_{2-x}$Sr$_{x}$CoIrO$_6$ compounds.\cite{Narayanan:10} In La$_2$CoIrO$_6$, neutron diffraction indicates E-type antiferromagnetic order with a distorted crystal structure ($P2_1/n$; monoclinic space group No.\,11), while for Sr$_{2}$CoIrO$_{6}$, A-type antiferromagnetism in a less distorted structure ($I2/m$; monoclinic space group No.\,12) is the most likely magnetic and crystalline structure. Visualizations of the crystal and magnetic structure of  La$_{2-x}$Sr$_{x}$CoIrO$_6$ are shown elsewhere.\cite{Narayanan:10} Due to the low neutron scattering cross section of Ir, the refinement does not include any useful information on the magnetic ordering at the Ir site. Due to the more strongly distorted structure in La$_2$CoIrO$_6$, a residual canted magnetic moment of the Co moments of about 1.65 $\mu_{\text{B}}$ per f.u. is obtained as evidenced by NPD measurements.\cite{Narayanan:10} Such a canted moment (or weak ferromagnetic behavior) does not occur in the opposite parent compound Sr$_2$CoIrO$_6$.

\begin{figure}[t]
\centering{
\includegraphics[width=0.9\columnwidth,clip=]{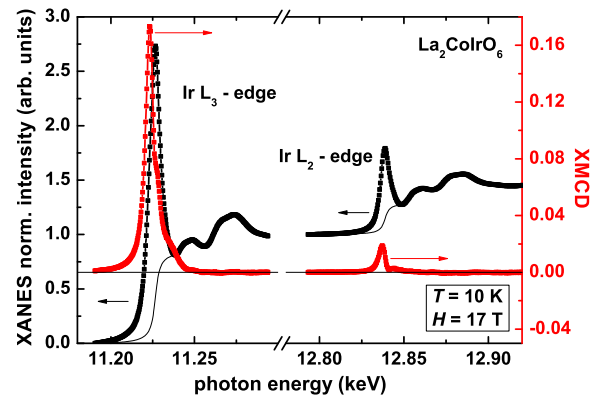}
 \caption{(Color online) XANES and XMCD normalized intensities at the $L_2$ and $L_3$ edges of Ir in La$_2$CoIrO$_6$. Units are arbitrary but can be compared in all figures.}\label{Fig:LCIO}}
\end{figure}

We present in Fig.~\ref{Fig:LCIO} the measurement of the XANES and XMCD signal of Ir in the end compound  La$_2$CoIrO$_6$ at 10 K and 17 T. A clear magnetic signal is detected, showing a substantial magnetization at the Ir site. Due to the high atomic mass of Ir, also orbital magnetism is expected to be substantial. Quantitatively, applying the standard sum rules, \cite{Thole:92,Carra:93} we derived a spin magnetic moment $m_{\text{spin}}=-0.205\,\mu_{\text{B}}$ and an orbital magnetic moment $m_{\text{orbital}}=-0.177\,\mu_{\text{B}}$ per Ir, resulting in a total magnetic moment $m_{\text{tot}}=-0.38\,\mu_{\text{B}}$ per Ir. Here, we have neglected the magnetic dipole contribution, thus, we can consider $m_{\text{spin}}$ as an effective spin magnetic moment. The result shows that the orbital contribution to the magnetic moment is of almost the same amount and sign as the spin contribution. Another key point is the {\em negative sign} of the Ir total magnetic moment. This unambiguously demonstrates the antiferromagnetic coupling of the Ir moment to the weak ferromagnetic moment/canted moment of the Co atoms.

\begin{figure}[t]
\centering{
\includegraphics[width=0.9\columnwidth,clip=]{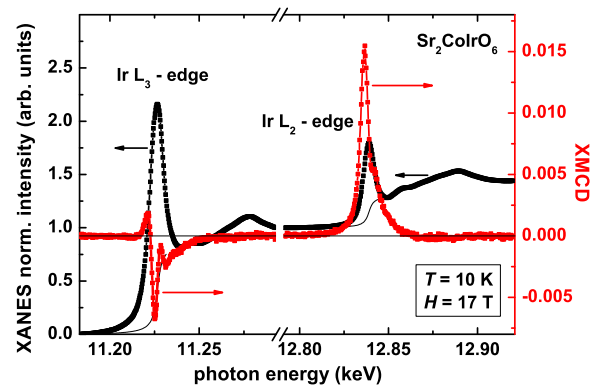}
 \caption{(Color online) XANES and XMCD normalized intensities at the $L_2$ and $L_3$ edges of Ir in Sr$_2$CoIrO$_6$.}\label{Fig:SCIO}}
\end{figure}

In Fig.~\ref{Fig:SCIO} we show the XANES and XMCD signal of Ir in the opposite parent compound  Sr$_2$CoIrO$_6$
at 10 K and 17 T. Here, we observe a completely different picture compared to La$_2$CoIrO$_6$.
Quantitatively, we derived a spin magnetic moment $m_{\text{spin}}=0.049\,\mu_{\text{B}}$ and an orbital magnetic moment $m_{\text{orbital}}=-0.01\,\mu_{\text{B}}$ per Ir, resulting in a total magnetic moment $m_{\text{tot}}=0.039\,\mu_{\text{B}}$ per Ir. This magnetic field induced moment is a paramagnetic moment aligned in the external field.

\begin{figure}[t]
\centering{
\includegraphics[width=0.9\columnwidth,clip=]{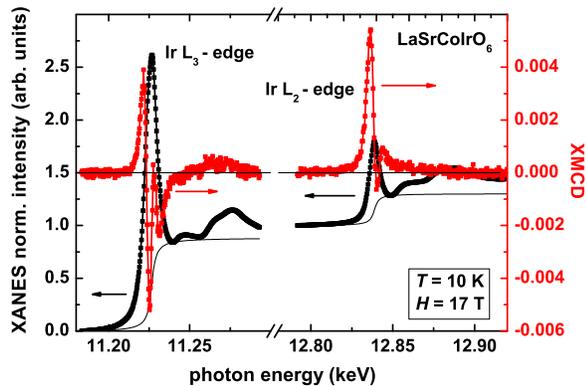}
 \caption{(Color online) XANES and XMCD normalized intensities at the $L_2$ and $L_3$ edges of Ir in LaSrCoIrO$_6$. }\label{Fig:LSCIO}}
\end{figure}

\begin{figure}[b]
\centering{
\includegraphics[width=0.9\columnwidth,clip=]{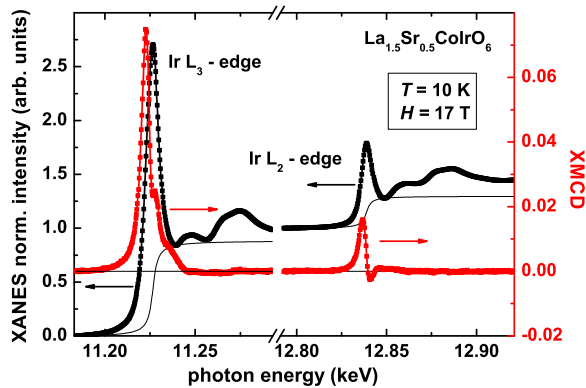}
 \caption{(Color online) XANES and XMCD normalized intensities at the $L_2$ and $L_3$ edges of Ir in La$_{1.5}$Sr$_{0.5}$CoIrO$_6$.}\label{Fig:L15S05CIO}}
\end{figure}

We also present the data on the intermediate compounds LaSrCoIrO$_6$ and La$_{1.5}$Sr$_{0.5}$CoIrO$_6$ in Fig.~\ref{Fig:LSCIO} and Fig.~\ref{Fig:L15S05CIO}, being aware of the fact that the crystal quality and homogeneity of such mixed compounds may be reduced compared to the parent compounds. In LaSrCoIrO$_6$, we obtain quantitatively a spin magnetic moment $m_{\text{spin}}=0.014\,\mu_{\text{B}}$ and an orbital magnetic moment $m_{\text{orbital}}=-0.003\,\mu_{\text{B}}$ per Ir, resulting in a total magnetic moment $m_{\text{tot}}=0.011\,\mu_{\text{B}}$ per Ir. This looks puzzling at first sight, since the total magnetic moment is close to 0. However, one has to note that going from one parent compound to the other, the total and spin magnetic moments change their sign. The compound LaSrCoIrO$_6$ seems to be close to the composition where the intrinsic behavior of Sr$_2$CoIrO$_6$ changes its character to an induced behavior as is most pronounced in La$_2$CoIrO$_6$.
La$_{1.5}$Sr$_{0.5}$CoIrO$_6$ clearly is in a transition state from La$_2$CoIrO$_6$ to LaSrCoIrO$_6$. For a better overview, we have summarized our quantitative results in Table~\ref{tab:table}.

\begin{table*}\caption{\label{tab:table} Summary of the Ir spin, orbital, and total magnetic moments (given in $\mu_{\text{B}}$/f.u.). $n_{5d}$ exp.~is the number of $d$-holes in Ir which was obtained experimentally by comparison to a Fe/Ir standard.\cite{Mansour:84,Wilhelm:01} We have used $n_{5d}$ exp.~for our calculations. For comparison, the theoretical value $n_{5d}$ th.~is shown, taken from band-structure calculations.\cite{Narayanan:10} $B$ is the branching ratio $L_3$/($L_3$+$L_2$), which is similar for all samples.}
\begin{ruledtabular}
\begin{tabular}{|l|r|r|r|r|r|r|r|}
  \hline
  compound        & $n_{5d}$ exp. & $n_{5d}$ th. & $m_{\text{spin}}$ & $m_{\text{orbital}}$ &  $\dfrac{m_{\text{orbital}}}{m_{\text{spin}}}$ & $m_{\text{tot}}$ & $B$ \\ \hline\hline
  Ir in Fe/Ir                     & 2.7  &      &        &        &        &        &       \\
  La$_2$CoIrO$_6$                 & 4.37 & 5.04 & -0.205 & -0.177 & 0.86   & -0.38  & 0.8   \\
  La$_{1.5}$Sr$_{0.5}$CoIrO$_6$   & 4.56 &      & -0.072 & -0.075 & 1.04   & -0.147 & 0.81  \\
  LaSrCoIrO$_6$                   & 4.63 & 5.23 & +0.014 & -0.003 & -0.193 & +0.011 & 0.80  \\
  Sr$_2$CoIrO$_6$                 & 4.11 & 5.37 & +0.049 & -0.01  & -0.197 & +0.039 & 0.78  \\
  \hline
\end{tabular}
\end{ruledtabular}
\end{table*}

\begin{figure}[t]
\centering{
\includegraphics[width=0.9\columnwidth,clip=]{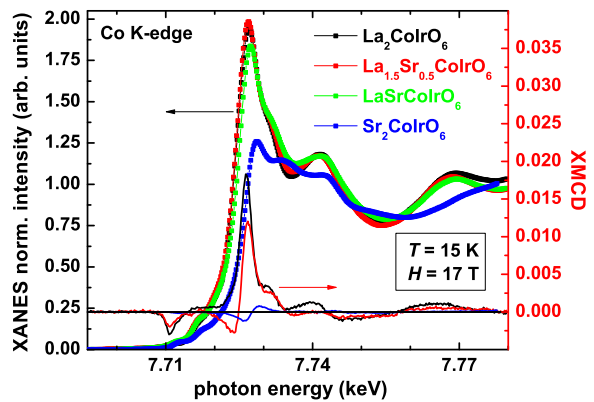}
 \caption{(Color online) XANES spectra for the Co $K$ edge in all compounds (left scale). The right scale shows the corresponding XMCD intensity.}\label{Fig:CoKedges}}
\end{figure}

\begin{figure}[b]
\centering{
\includegraphics[width=0.9\columnwidth,clip=]{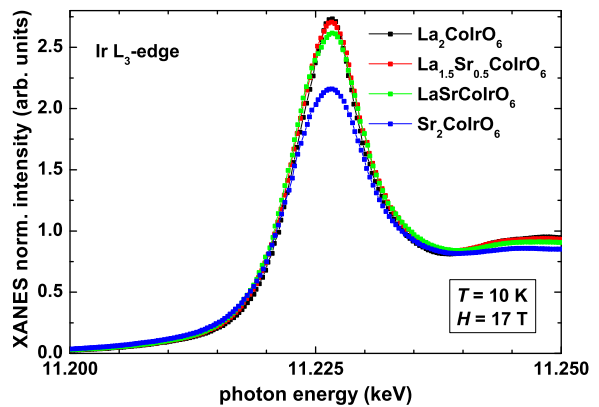}
 \caption{(Color online) Comparison of the XANES spectra of the Ir $L_3$ edge in all compounds.}\label{Fig:Iredges}}
\end{figure}

Figures \ref{Fig:CoKedges} and \ref{Fig:Iredges} show the Co $K$-edge and Ir $L_3$-edge XAS spectra for all samples for a better comparison. There is a clear shift from the Co white line of La$_2$CoIrO$_6$ to Sr$_2$CoIrO$_6$ indicating the transition from Co$^{2+}$ to Co$^{3+}$. For the Ir spectra, in contrast, such a shift is not clearly observable. Furthermore, the intensity of the white line of Ir in Sr$_2$CoIrO$_6$ is lower than for La$_2$CoIrO$_6$, which is unexpected. However, also in other cases the Ir white line intensity and position is not very much shifted or changed as a function of the Ir valence state.\cite{Mugavero:09} The reason is most likely the fact that $5d$ transition metals have much more diffuse valence orbitals compared to $3d$ transition metals.

\begin{figure}[b]
\centering{
\includegraphics[width=0.9\columnwidth,clip=]{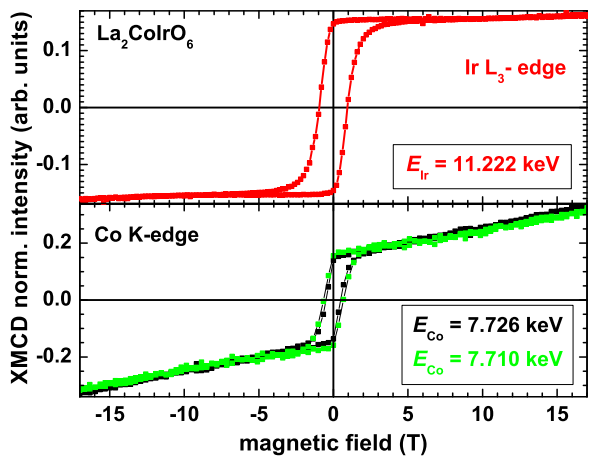}
 \caption{(Color online) XMCD hysteresis curves of Ir $L_3$ edge and Co $K$ edge in La$_2$CoIrO$_6$. The Co hysteresis has been measured at two different energies corresponding to the pronounced XMCD features at the pre-edge and at the edge shown in Figure~\ref{Fig:CoKedges}.}\label{Fig:hyster}}
\end{figure}

In Fig.~\ref{Fig:hyster}, the element specific magnetization curves of La$_2$CoIrO$_6$ recorded by monitoring the Ir $L_3$-edge XMCD and the Co $K$-edge XMCD signal as a function of applied magnetic field are shown. The temperature was calibrated by measurement of the hysteresis loops as a function of temperature in a SQUID. The Co magnetization curves have been recorded at two XMCD values: at the pre-peak feature (7710.64\,eV) and at the edge (7726.35\,eV). For both energies, the same behavior is observed. A striking feature is the strong linear contribution beyond the hysteresis loop. This behavior is due to the continuous field alignment of the canted Co magnetic moments. Note that the Co and Ir moments are strongly coupled to each other. The Ir magnetization also shows a hysteresis loop, with the magnetic moment of Ir coupled negatively to the Co moment. Furthermore, there is a slight increase in the Ir moment with increasing field following the Co magnetization, however, with a 15 times smaller slope. This shows that the direct  exchange coupling and the dipolar coupling, which are both proportional to the magnetic moment of Co and Ir, are small, and gives evidence that the magnetization of Ir is related to the hybridization mechanism as described in the next paragraph. In contrast to La$_2$CoIrO$_6$, the magnetic hysteresis for Sr$_2$CoIrO$_6$ (not displayed) shows for both edges an almost perfectly linear behavior as expected for an antiferromagnet or a paramagnet.

\begin{figure}[t]
\centering{
\includegraphics[width=0.95\columnwidth,clip=]{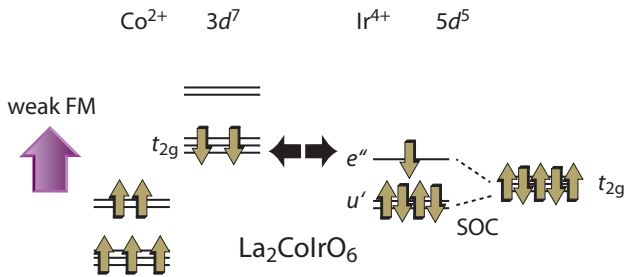}
 \caption{(Color online) Model of magnetic coupling between Co and Ir in La$_{2}$CoIrO$_{6}$: Hybridization of Co $t_{2\text{g}}$ spin-down levels with the Ir $e''$ level leads to an induced negative moment on the Ir site. The splitting of the Ir $t_{2\text{g}}$ levels is due to the spin-orbit coupling (SOC).}\label{Fig:model}}
\end{figure}

We suggest a simple model including spin-orbit coupling to explain the magnetic coupling in the compounds La$_2$CoIrO$_6$ and Sr$_2$CoIrO$_6$. In the case of La$_2$CoIrO$_6$ we are dealing with a Co$^{2+}$ $3d^7$ and Ir$^{4+}$ $5d^5$ combination. Assuming a strong spin splitting according to Hund's rule and a crystal field splitting, we find Co spin-down electrons in the $t_{2\text{g}}$ orbital at the Fermi surface. At the Ir site, we have a strong crystal field splitting, but almost no spin splitting, leaving an equal amount of spin-up and spin-down electrons at the Fermi level in the first step. The spin-orbit coupling splits the $t_{2\text{g}}$ level in one fully occupied $u'$ level and one single occupied $e''$ level. Switching on the hybridization between the spin-down $t_{2\text{g}}$ orbitals of Co and the $e''$ level of Ir, a kinetic energy gain can only be obtained by spin-down electrons due to the strong Hund's coupling at the Co site. This will create a tendency to accumulate spin-down electrons at the Ir$^{4+}$ site, explaining naturally the observed {\em negative} magnetization of Ir. In this model, the residual weak ferromagnetic moment of Co couples antiferromagnetically to the Ir moment, which is, at the same time, enhanced by the described hybridization. The suggested coupling scheme explaining the experimentally observed main features is schematically shown in Fig.~\ref{Fig:model}.

Let us now consider the case of Sr$_2$CoIrO$_6$, where we are dealing with a Co$^{3+}$ $3d^6$ and Ir$^{5+}$ $5d^4$ combination. From neutron diffraction we know that Co orders antiferromagnetically without canting. Due to the antiferromagnetic order of the Co ions, hybridization cannot induce a spin imbalance at the Ir site. Furthermore, the Ir spin-orbit coupling leads to a fully occupied $u''$ level, hampering further  hybridization. Therefore, the residual moment on the Ir site is an intrinsic paramagnetic moment. In consequence, this Ir moment aligns with the external field as observed. Comparing the absolute values of the SQUID data to the XMCD results, the magnetic moment in Sr$_2$CoIrO$_6$ originates almost exclusively from the Ir effective number of magnetons, which is about 0.47 $\mu_{\text{B}}$. In a simple ionic picture as measured in Ir complexes, the paramagnetic moment of Ir is of the order of 1.4 to 1.8 $\mu_{\text{B}}$/Ir.\cite{Sloth:54,Norman:59} The reduced paramagnetic moment observed in the double perovskite is obviously due to bandstructure effects.

\begin{table}\caption{\label{tab:table2} Summary of total Ir moments $m_{\text{tot}}$ measured at temperature $T$, in different compounds with Curie-temperature $T_{\text{C}}$. Paramagnetic moments are marked as para.}
\begin{ruledtabular}
\begin{tabular}{|l|r|r|r|}
  \hline
  compound        & $\mu_{\text{tot}}$ [$m_{\text{B}}$/Ir] & $T$ [K] & $T_{\text{C}} [K]$ \\ \hline\hline
  IrMnAl  \cite{Krishnamurthy:03a}     & 0.015  &  30  & 379  \\
  IrMnAl  \cite{Krishnamurthy:03a}     & 0.0055 &  297 & 379  \\
  Fe$_2$IrSi  \cite{Krishnamurthy:03b} & 0.15   &  297 & 662  \\
  Sr$_2$IrO$_4$ \cite{Kim:09}          & 0.075  &  5   & 240  \\
  BaIrO$_3$  \cite{Cao:04}             & 0.03   &  5   & 175  \\
  BaIrO$_3$  \cite{Cao:04}             & 0.13   & para & 175 \\ \hline
  La$_2$CoIrO$_6$  (this paper)        & -0.38  & 10   & 90   \\
  Sr$_2$CoIrO$_6$  (this paper)        & 0.47   & para & --    \\
  \hline
\end{tabular}
\end{ruledtabular}
\end{table}

In Table~\ref{tab:table2} we compare the measured values of the total Ir magnetization $m_{\text{tot}}$ in different compounds. While the paramagnetic moments are well below the expected values for the ionic $S=1/2$ picture, but largest in Sr$_2$CoIrO$_6$, ordered magnetic moments are low in the absence of strongly magnetic ions (Mn only shows a relatively small magnetic moment in IrMnAl). Compared to IrMnAl, in the Heusler alloy Fe$_2$IrSi \cite{Krishnamurthy:03b} the (spin) magnetic moment of Ir is 10 times larger, but still 3 times smaller than in La$_2$CoIrO$_6$. In both cases, the hybridization with highly spin-polarized orbitals leads to an enhanced magnetic moment of Ir. In the case of La$_2$CoIrO$_6$ the large orbital magnetic moment of Ir indicates a strong spin-orbit coupling which is three times larger than that of Fe$_2$IrSi.

Recently, interest has arisen in Ir-based oxide compounds due to the observation of a spin-orbital Mott state with $J=1/2$ in Sr$_2$IrO$_4$ \cite{Kim:09} and a huge Ir $5d$ orbital moment, larger than the spin moment, in BaIrO$_3$.\cite{Laguna-Marco:10} The effective $J=1/2$ state originates from the spin-orbit interaction and the single occupied $e'$ level, which also plays a key role in our hybridization picture. However, although in La$_2$CoIrO$_6$ also the orbital magnetic moment is comparable to the spin moment, the coupling mechanisms are completely different in the double-perovskite compounds compared to the layered structures of Sr$_2$IrO$_4$ and BaIrO$_3$. In these two compounds, the magnetic interaction is dominated by the Ir-Ir interactions. Furthermore, both compounds have strong two- and one-dimensional structural characteristics.\cite{Cao:04} In contrast, in the three-dimensional double perovskites discussed here, the magnetic interaction is dominated by the strongly magnetic ion Co and the Co-Ir interaction. In BaIrO$_3$ and Sr$_2$IrO$_4$, Ir$^{4+}$ has the strongly reduced magnetic moment of 0.03\,$\mu_{\text{B}}$/Ir resp.~0.075\,$\mu_{\text{B}}$/Ir, which is 13 resp.~5 times smaller than in La$_2$CoIrO$_6$, again underlining the relevance of the mechanism of hybridization induced magnetic moments in Ir in the double perovskite structure suggested here. This mechanism explains our experimental observation that Ir in La$_2$CoIrO$_6$ has the highest ordered magnetic moment reported so far.

\section{\textbf{Summary}}

We have studied the iridium magnetism in the double perovskite structure of
the antiferromagnetic resp.~weakly ferromagnetic compounds La$_{2-x}$Sr$_x$CoIrO$_6$. In the case of a weak ferromagnetic moment of canted Co spins and strong spin-orbit coupling, we have shown that Ir couples antiferromagnetically to the residual Co moment, while at the same time, an unusually large magnetic moment is induced at the Ir site. In the case of perfect antiferromagnetic order of Co, the Ir ions possess a paramagnetic moment. In total, our results show that the heavy ion Ir is susceptible to the induction of considerable ordered magnetic moments in the double perovskite structure.

\section*{\textbf{Acknowledgement}}
This work was supported by the LOEWE-Centre AdRIA, by DFG through Grant No. SPP 1178,
and by the European Synchrotron Radiation Facility (Grant Nos. HE-2379, HE-2848, and HE-3567).

\end{document}